\documentclass[12pt]{iopart}
\usepackage{graphicx} 
\usepackage{cite} 

\begin{document}

\title{Remote Cross-resonance Gate between Superconducting Fixed-frequency Qubits}
\author{Mari Ohfuchi and Shintaro Sato}
\address{Quantum Laboratory, Fujitsu Research, Fujitsu Limited, Atsugi 243-0197, Japan}
\ead{mari.ohfuti@fujitsu.com}
\vspace{10pt}
\begin{indented}
\item 17 December 2023
\end{indented}

\begin{abstract}
High-fidelity quantum state transfer and remote entanglement between superconducting fixed-frequency qubits have not yet been realized. In this study, we propose an alternative remote cross-resonance gate. Considering multiple modes of a superconducting coaxial cable connecting qubits, we must find conditions under which the cross-resonance gate operates with a certain accuracy even in the presence of qubit frequency shifts due to manufacturing errors. For 0.25- and 0.5-m cables, remote cross-resonance gates with a concurrence of $>99.9\%$ in entanglement generation are obtained even with $\pm$10-MHz frequency shifts. For a 1-m cable with a narrow mode spacing, a concurrence of 99.5\% is achieved by reducing the coupling between the qubits and cable. The optimized echoed raised-cosine pulse duration is 150--400 ns, which is similar to the operation time of cross-resonance gates between neighboring qubits on a chip. The dissipation through the cable modes does not considerably affect the obtained results. Such high-precision quantum interconnects pave the way not only for scaling up quantum computer systems but also for nonlocal connections on a chip.

\end{abstract}

\maketitle

\section{\label{sec:introduction}Introduction}

\begin{figure}[h]
\centering
\includegraphics[width=80mm]{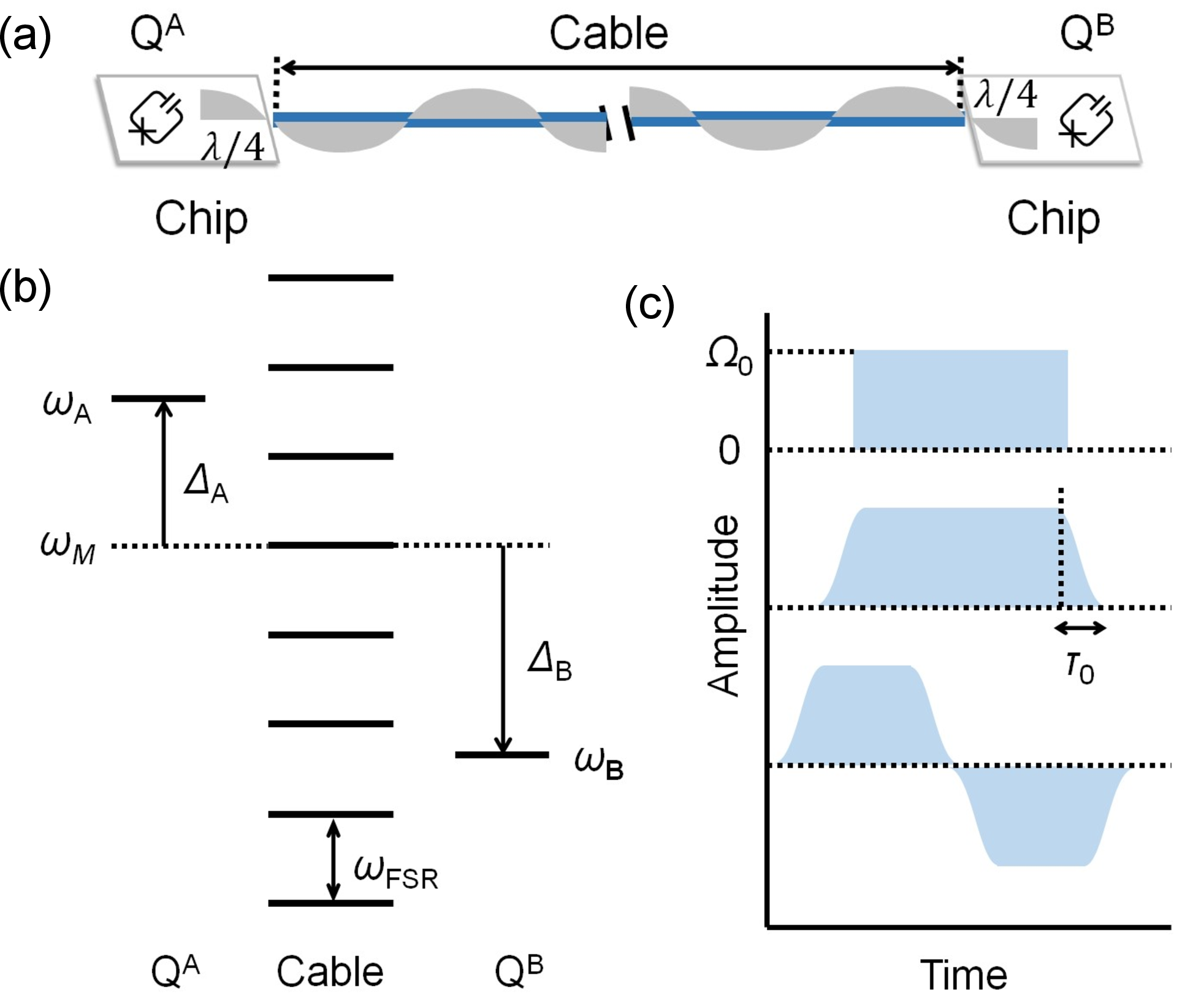}
\caption{\label{fig:concept}(a) Schematic of quantum interconnects via a coaxial cable and coplanar waveguides, where $\mathrm{Q^A}$ and $\mathrm{Q^B}$ are qubits and $\lambda$ represents the wavelength. (b) Energy diagram: $\omega_\mathrm{A}$, $\omega_\mathrm{B}$, and $\omega_M$ are the frequencies of $\mathrm{Q^A}$, $\mathrm{Q^B}$, and the $M$th mode of the cable, respectively. The detunings of $\omega_\mathrm{A}$ and $\omega_\mathrm{B}$ measured from $\omega_M$ are denoted by ${\it{\Delta}}_\mathrm{A}$ and ${\it{\Delta}}_\mathrm{B}$, respectively. The free spectral range $\omega_\mathrm{FSR}$ corresponds to the equally spaced frequencies of the cable mode. (c) Schematic of cross-resonance pulse amplitudes: flat, raised-cosine, and echoed pulses from the top to bottom. The amplitude of the flat parts is $\Omega_0$, and the rise and fall times are $\tau_0$.}
\end{figure}

Superconducting quantum computers are a promising candidate for realizing practical large-scale quantum computation in the future \cite{Byrd2023,Bravyi2022}. To scale up quantum computer systems, different levels of modularity and interconnects between them have been actively investigated 
\cite{Bravyi2022,PRXQ2021,Gold2021,Niu2023,Zhong2021,Mirhosseini2020}. 
One approach is to connect adjacent quantum chips with sufficiently short distances ($\sim$1 mm), which are the same as the distance between qubits on a single chip \cite{Gold2021,Conner2021,Kosen2022,Marxer2023}, practically increasing the chip size and possibly allowing mapping of logical qubits encoded using surface codes \cite{Campbell2017,Bravyi1998,Dennis2002,Kitaev2003,Raussendorf2007,Fowler2012,Horsman2012,Zhao2022,Krinner2022,Google2023}. Quantum chips separated by long distances in a refrigerator are connected through superconducting coaxial cables that pass microwave signals \cite{Niu2023,Zhong2021,Cirac1997,Kurpiers2018,Axline2018,Campagne2018,Leung2019,Zhong2019,Magnard2020,Chang2020,Burkhart2021}, millimeter-wave photonic links \cite{Pechal2017}, or acoustic transmission lines as quantum phononic channels \cite{Safavi-Naeini2019,Dumur2021}. For connecting qubits in different refrigerators or sending quantum information to the so-called quantum internet, quantum information stored in superconducting qubits must be frequency-converted into optical photons \cite{Mirhosseini2020,Han2021,Andrews2014,Williamson2014,Hisatomi2016,Krastanov2021,Tu2022,Honl2022,Chiappina2023}. However, it is very challenging to overcome large energy differences and achieve high conversion efficiencies. Herein, we focus on interconnects via 0.1--1-m superconducting coaxial cables inside a refrigerator [Fig. \ref{fig:concept}(a)]. Such medium-range quantum interconnects pave the way not only for scaling up quantum computer systems but also for nonlocal connections on a chip and thus non-two-dimensional (non-2D) quantum error-correcting codes \cite{Bravyi2022,Campbell2017}.

Quantum state transfer between two qubits through a superconducting coaxial cable has been investigated using the pitch-and-catch scheme, wherein a photon emitted by a qubit at one end of the communication channel is received by another qubit at the opposite end \cite{PRXQ2021,Cirac1997,Kurpiers2018,Axline2018,Campagne2018}. In 2018, a fidelity of 80\% was reported for quantum state transfer using a 0.9-m coaxial cable \cite{Kurpiers2018}. Half-quantum state transfer can generate a remote entangled state. A fidelity of 79\% was reported for half-quantum state transfer \cite{Kurpiers2018}. Recent advances in fidelities have been achieved through a quantum state transfer scheme using the standing-wave modes of cables \cite{Niu2023,Zhong2021,Leung2019,Zhong2019}. A fidelity of 91\% was reported for both quantum state transfer and remote entanglement through a 1-m cable \cite{Zhong2021}. Although the cable length was 0.25 m, the fidelity was increased to 99\%  by employing an aluminum low-loss coaxial cable and minimizing the dissipation at the junction between the cable and coplanar waveguides [Fig. \ref{fig:concept}(a)] \cite{Niu2023}. However, the settings for these experiments are limited to the combination of frequency-tunable qubits \cite{Barends2014,DiCarlo2009,Kelly2015,Reagor2018} and tunable couplers \cite{Chen2014} for matching the qubit frequency to the standing-wave mode of the cable. This can be because a small frequency detuning substantially reduces the transfer efficiency (\ref{A:transfer}).

Fixed-frequency qubits have the advantages of long coherence times and noise immunity \cite{Koch2007,Allman2014,Sirois2015,Axline2016,Reagor2016,McKay2016,Lu2017,Jurcevic2021,Place2021}. Low sensitivity to low-frequency charge and magnetic-flux fluctuations ensures qubit coherence. However, with current manufacturing processes of fixed-frequency qubits, slight frequency shifts are unavoidable \cite{Hertzberg2021,Zhang2022}. Even if qubits are fabricated at a frequency that matches the standing-wave mode of the cable, as mentioned above, a very small frequency detuning of 5 MHz [0.1\% of a typical qubit frequency (approximately 5 GHz)] severely compromises the transfer efficiency. Thus, quantum state transfer using cable standing-wave modes is not applicable to fixed-frequency qubits. A two-qubit gate for fixed-frequency qubits is well known to use the cross-resonance effect \cite{Rigetti2010,Chow2011,Corcoles2013,Sheldon2016,Patterson2019,Sundaresan2020,Kandala2021,Heya2021,Malekakhlagh2022}. Cross-resonance microwave signals enable conditional control between coupled qubits of different frequencies. Recent improvements in control pulses and coupling circuits have increased the fidelity of cross-resonance gates to 99.7\%, with gate times of 100--400 ns \cite{Sheldon2016,Sundaresan2020,Kandala2021}. In this study, we consider cross-resonance gates between two fixed-frequency qubits connected by a coaxial cable. More specifically, we explore the possibility of cross-resonance gates acting through multiple cable modes with an energy spacing similar to or narrower than qubit frequency detunings suitable for the cross-resonance effect.

The remainder of this paper is organized as follows. In Sec. II, we present the configuration of the transmission paths and reveal the fundamental properties of remote cross-resonance gates, such as concurrences as measures of entanglement generation and average gate fidelities. Section III focuses on the dissipative property of the transmission paths and its effect on the cross-resonance gates. In Sec. IV, we discuss leakage and single-qubit gate properties in remote cross-resonance gate settings. Finally, Sec. V presents the conclusions of the study and an outlook for future works.  

\section{\label{sec:rcr}Remote cross-resonance gate}

We first configure the transmission paths and then present the fundamental properties of remote cross-resonance gates via these transmission paths. The qubit frequency dependence of the concurrence in entanglement generation is an important result of this work. We also introduce the average gate fidelity as a more general characteristic of two-qubit gates. In this section, we do not consider any dissipation to focus on the effect of multiple cable modes.

\subsection{Transmission path}

We consider a transmission path comprising a cable and coplanar waveguides as shown in Fig. \ref{fig:concept}(a). Target cable lengths are 0.25, 0.5, and 1 m. Let the frequency of the $M$th standing-wave mode of the cable be $\omega_M/(2\pi)=5$ GHz [Fig. \ref{fig:concept}(b)]. Here, $M$ is an odd number and is chosen such that the cable length is close to the target cable length. The cable and coplanar waveguide lengths are determined as follows:
\begin{eqnarray}
l_\mathrm{Cable} &=& \frac{M-1}{2}\frac{v_\mathrm{Cable}}{\omega_M/(2\pi)}=\frac{M-1}{2} \lambda_\mathrm{Cable}\nonumber\\
l_\mathrm{CPW} &=& \frac{1}{4}\frac{v_\mathrm{CPW}}{\omega_M/(2\pi)}=\frac{1}{4} \lambda_\mathrm{CPW},
\end{eqnarray}
where the microwave speeds in the cable and coplanar waveguide, $v_\mathrm{Cable}=2.472\times10^8$ m/s and $v_\mathrm{CPW}=1.157 \times10^8$ m/s, respectively, were determined to reproduce the experimental results \cite{Niu2023} and $\lambda_\mathrm{Cable}$ and $\lambda_\mathrm{CPW}$ are the microwave wavelengths in the cable and coplanar waveguide, respectively. The standing-wave mode spacing, namely free spectral range, is obtained as $\omega_\mathrm{FSR}=\omega_M/M$.
These characteristics of transmission paths are summarized in Table \ref{table:path}. 
\begin{table}[tbh]
\caption{\label{table:path}Cable length ($l_\mathrm{Cable}$), coplanar waveguide length ($l_\mathrm{CPW}$), number of cable standing-wave mode for 5 GHz ($M$), and free spectral range ($\omega_\mathrm{FSR}$) for target cable lengths of 0.25, 0.5, and 1 m.}
\centering
\begin{tabular}{@{}cccccccc}
\br
 $l_\mathrm{Cable}$ (m) & $l_\mathrm{CPW}$ (mm) & $M$ & $\omega_\mathrm{FSR}/(2\pi)$ (GHz)\\
\mr
0.2966 & 5.8 & 13 & 0.3846\\
0.5438 & 5.8 & 23 & 0.2174\\
1.0832 & 5.8 & 43 & 0.1163\\
\br
\end{tabular}
\end{table}

\subsection{Remote cross-resonance Hamiltonian}

Two fixed-frequency qubits $\mathrm{Q^A}$ and $\mathrm{Q^B}$ are connected to the transmission path [Fig. \ref{fig:concept}(a)]. The Hamiltonian for remote cross-resonance gates is given by
\begin{eqnarray}
\label{Hamiltonian}
H_0 &=& {\it{\Delta}}_\mathrm{A}\sigma_\mathrm{A}^{\dagger}\sigma_\mathrm{A} + {\it{\Delta}}_\mathrm{B}\sigma_\mathrm{B}^{\dagger}\sigma_\mathrm{B}\nonumber\\
&+& \sum_{m=M_\mathrm{min}}^{M_\mathrm{max}}(m-M)\omega_\mathrm{FSR}\sigma_m^{\dagger}\sigma_m\nonumber\\
&+& \sum_{m=M_\mathrm{min}}^{M_\mathrm{max}} g_\mathrm{A}\sqrt{\omega_m/\omega_M}(\sigma_\mathrm{A}\sigma_m^{\dagger}+\sigma_\mathrm{A}^{\dagger}\sigma_m)\nonumber\\
&+& \sum_{m=M_\mathrm{min}}^{M_\mathrm{max}}(-1)^m g_\mathrm{B}\sqrt{\omega_m/\omega_M} (\sigma_\mathrm{B}\sigma_m^{\dagger}+\sigma_\mathrm{B}^{\dagger}\sigma_m)\nonumber\\
H &=& H_0+\Omega(t)\cos(\tilde{\omega}_\mathrm{B}t)\sigma_x^\mathrm{A},
\end{eqnarray}
where  $\sigma_\mathrm{A}$, $\sigma_\mathrm{B}$, and  $\sigma_m$ are annihilation operators for $\mathrm{Q^A}$, $\mathrm{Q^B}$, and the $m$th cable mode, ${\it{\Delta}}_\mathrm{A}$ and ${\it{\Delta}}_\mathrm{B}$ are the qubit frequency detunings measured from $\omega_M$, and $\omega_m=m\omega_\mathrm{FSR}$ is the $m$th cable mode frequency. The couplings between the qubits and the $M$th cable mode are $g_\mathrm{A}$ and $g_\mathrm{B}$ and proportional to the square root of the frequency in multimode coupling \cite{Zhong2021,Sundaresan2015}. 

The last term of the second equation in Eq. (\ref{Hamiltonian}) represents a cross-resonance effect, and $\tilde{\omega}_\mathrm{B}$ is the dressed qubit frequency of $\mathrm{Q^B}$ that is by solving an eigenvalue problem for $H_0$. The cross-resonance pulse envelope $\Omega(t)$ is schematically presented in Fig. \ref{fig:concept}(c). We consider three pulse envelopes: flat pulse envelope, raised-cosine pulse envelope to reduce spectral leakage in the frequency domain \cite{Motzoi2009,Gambetta2011}, and echoed pulse envelope to equalize the di?erence in pulse duration due to the initial state \cite{Corcoles2013,Sheldon2016}. We fix the amplitude of the ?at parts to $\Omega_0/(2\pi)=0.1\ \mathrm{GHz}$ and the rise and fall times for the raised-cosine and echoed pulses to $\tau_0=100\times2\pi/\tilde{\omega}_\mathrm{B} \sim20\ \mathrm{ns}$ because the cross-resonance effect can be adjusted by the total pulse duration $\tau$.

We write the qubit states as $|{\it{\Phi}}_i\rangle=c_{i,|00\rangle}|00\rangle+c_{i,|01\rangle}|01\rangle+c_{i,|10\rangle}|10\rangle+c_{i,|11\rangle}|11\rangle$, where each ket represents the eigenstate of $H_0$ in order of $\mathrm{Q^A Q^B}$. 
The ideal cross-resonance gate is expressed as
\begin{equation}
\label{rcr}
[ZX]^{1/2} = \frac{1}{\sqrt{2}}
\left( \begin{array}{cccc}
1 & -i & 0 & 0 \\
-i & 1 & 0 & 0 \\
0 & 0 & 1 & i \\
0 & 0 & i & 1 \\
\end{array} \right),
\end{equation}
which is combined with two single-qubit gates to form a CNOT gate (i.e., $\mathrm{CNOT}=[ZI]^{-1/2}[ZX]^{1/2}[IX]^{-1/2}$). The CNOT gate results in remote state transfer for the initial state of $|{\it{\Phi}}^\mathrm{I}_1\rangle=|10\rangle$ and forms a Bell state as a maximum entangled state for $|{\it{\Phi}}^\mathrm{I}_+\rangle=(|00\rangle+|10\rangle)/\sqrt{2}$.  

\subsection{Time evolution simulation}

\begin{figure}[tbh]
\centering
\includegraphics[width=80mm]{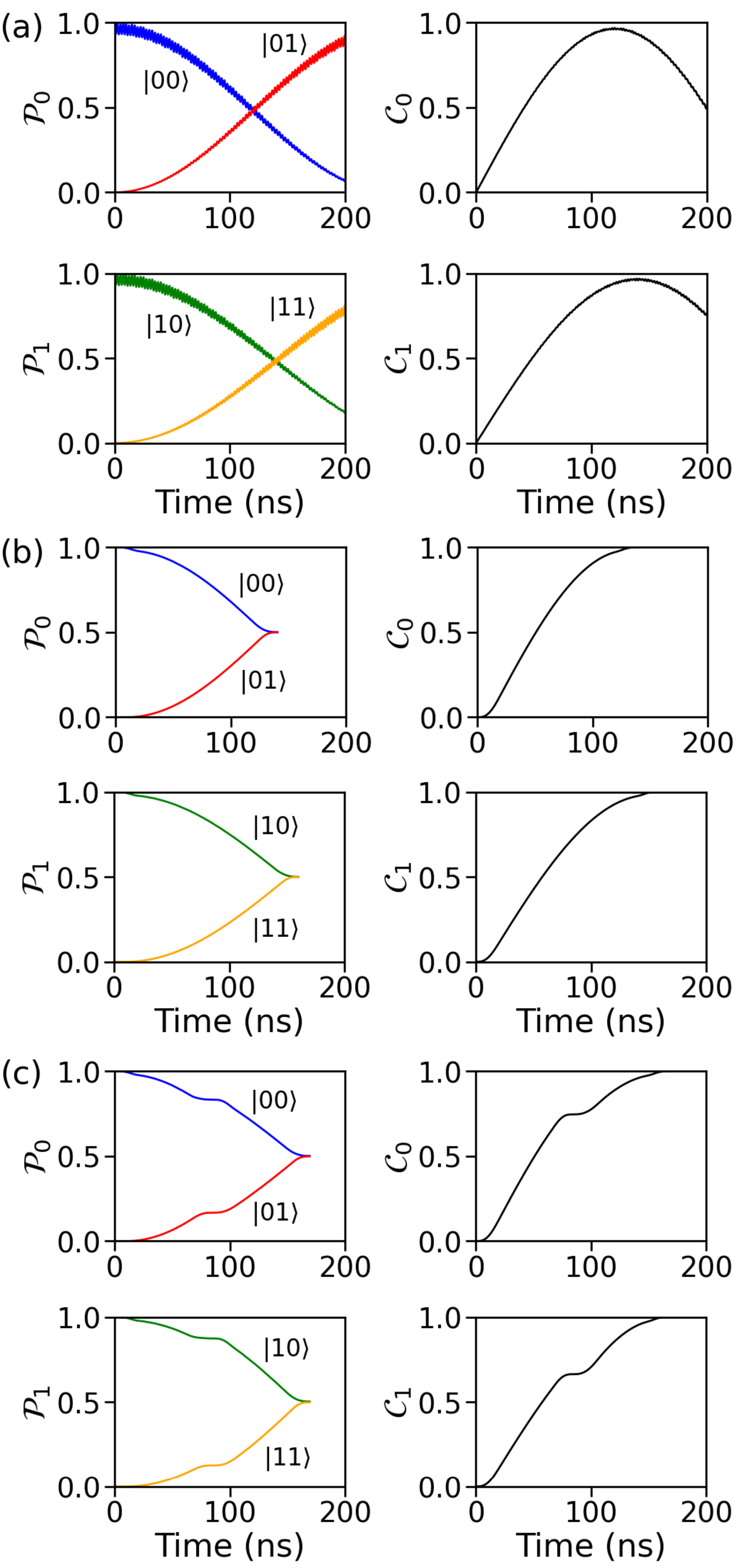}
\caption{\label{fig:025m} Time evolution with (a) flat, (b) raised-cosine, and (c) echoed cross-resonance pulses for a 0.25-m coaxial cable: occupation probability of the qubit states $\mathcal{P}$ and concurrence $\mathcal{C}$. For each figure, the upper and lower panels show the results with the initial qubit states of $|{\it{\Phi}}^\mathrm{I}_0\rangle=|00\rangle$ and $|{\it{\Phi}}^\mathrm{I}_1\rangle=|10\rangle$, respectively. Other parameters are $g/(2\pi)=0.03$ GHz, $\delta_\mathrm{A}=0.7$, and $\delta_\mathrm{B}=0.3$ [${\it{\Delta}}/(2\pi)=0.3846$ GHz].}
\end{figure}

Herein, we present details of time evolution simulations using the case of a 0.25-m coaxial cable. In the Hamiltonian of Eq. (\ref{Hamiltonian}), we set $M=13$ and $\omega_\mathrm{FSR}/(2\pi)=0.3846$ GHz from Table \ref{table:path}. Let ${\it{\Delta}}_\mathrm{A}=0.7\omega_\mathrm{FSR}$ and ${\it{\Delta}}_\mathrm{B}=-0.3\omega_\mathrm{FSR}$, then $\omega_\mathrm{A}>\omega_M$ and $\omega_\mathrm{B}<\omega_M$ as shown in Fig. \ref{fig:concept}(b). This is written as $\delta_\mathrm{A}=|{\it{\Delta}}_\mathrm{A}|/\omega_\mathrm{FSR}=0.7$ and $\delta_\mathrm{B}=|{\it{\Delta}}_\mathrm{B}|/\omega_\mathrm{FSR}=0.3$ and results in a qubit frequency detuning of ${\it{\Delta}}/(2\pi)=(\omega_\mathrm{A}-\omega_\mathrm{B})/(2\pi)=0.3846\ \mathrm{GHz}$. Two cable modes above $\omega_\mathrm{A}$ and two below $\omega_\mathrm{B}$ yield the upper and lower cable mode limits $M_\mathrm{max}=15$ and $M_\mathrm{min}=11$, respectively. The coupling between the qubits and cable modes is chosen as $g_\mathrm{A}/(2\pi)=g_\mathrm{B}/(2\pi)=g/(2\pi)=0.03$ GHz. We numerically simulate time evolution using the QuTip framework \cite{Johansson2012,Johansson2013}. The results with a cross-resonance pulse are presented in Figure \ref{fig:025m}.

We first consider the initial qubit states of $|{\it{\Phi}}^\mathrm{I}_0\rangle=|00\rangle$ and  $|{\it{\Phi}}^\mathrm{I}_1\rangle=|10\rangle$. When $[ZX]^{1/2}$ is applied,  $|{\it{\Phi}}^\mathrm{I}_0\rangle$ and $|{\it{\Phi}}^\mathrm{I}_1\rangle$ become $|{\it{\Phi}}^\mathrm{F}_0\rangle=(|00\rangle-i|01\rangle)/\sqrt{2}$ and $|{\it{\Phi}}^\mathrm{F}_1\rangle=(|10\rangle+i|11\rangle)/\sqrt{2}$, respectively.
Figure \ref{fig:025m}(a) presents the results with a flat cross-resonance pulse applied continuously for 200 ns. The amplitude of the cross-resonance pulse is set to $\Omega_0/(2\pi)=0.1\ \mathrm{GHz}$, as described above. The occupation probability of the qubit state is given as $\mathcal{P}_{i,k}=|c_{i,k}|^2$ for each $k$ state. The $\mathcal{P}_0$ for $|{\it{\Phi}}^\mathrm{I}_0\rangle$ and $\mathcal{P}_1$ for $|{\it{\Phi}}^\mathrm{I}_1\rangle$ are shown in the left panels. With increasing time, $\mathcal{P}_{0,|00\rangle}$ decreases from 1, $\mathcal{P}_{0,|01\rangle}$ increases from 0, and $\mathcal{P}_{0,|00\rangle}$ and $\mathcal{P}_{0,|01\rangle}$ intersect at approximately $\mathcal{P}_0=0.5$ and 120 ns. Similarly, $\mathcal{P}_{1,|10\rangle}$ and $\mathcal{P}_{1,|11\rangle}$ intersect at approximately $\mathcal{P}_1=0.5$ and 140 ns.
We define the concurrences
\begin{eqnarray}
\label{concurrence}
\mathcal{C}_0 & = & 2|c_{0,|00\rangle}||c_{0,|01\rangle}| \nonumber\\
\mathcal{C}_1 & = & 2|c_{1,|10\rangle}||c_{1,|11\rangle}|
\end{eqnarray}
so that they become 1 when $|{\it{\Phi}}_i\rangle=|{\it{\Phi}}^\mathrm{F}_i\rangle$. The right panels of Fig. \ref{fig:025m}(a) show $\mathcal{C}_0$ and $\mathcal{C}_1$. The concurrence takes the maximum values of $\mathcal{C}_0=96.82\%$ at 120.1 ns and $\mathcal{C}_1=96.86\%$ at 139.4 ns, which are near the time of the intersection of each $\mathcal{P}$. 
These results show that the cross-resonance pulses with parameters currently under consideration can fundamentally act as a $[ZX]^{1/2}$ gate with a pulse duration similar to those of cross-resonance gates between neighboring qubits on a chip \cite{Sheldon2016,Kandala2021}.

To reduce spectral leakage in the frequency domain, a raised-cosine pulse is employed. We fix the rise and fall time to $\tau_0=100\times 2\pi/\tilde{\omega}_\mathrm{B}=20.5\ \mathrm{ns}$ and find the pulse duration $\tau$ when $\mathcal{C}_0$ and $\mathcal{C}_1$ take their maximum values. We obtain the maximum value of $\mathcal{C}_0=\mathcal{C}_1=100.00\%$ at $\tau=140.4\ \mathrm{ns}$ for $|{\it{\Phi}}^\mathrm{I}_0\rangle$ and  $\tau=159.9\ \mathrm{ns}$ for $|{\it{\Phi}}^\mathrm{I}_1\rangle$. The time evolution during the raised-cosine pulse for each initial state is depicted in Fig. \ref{fig:025m}(b). 

Further, to equalize the difference in $\tau$ due to the initial state, we adopt an echoed cross-resonance pulse. To isolate the precision of remote cross-resonance gates, two single-qubit $\pi$ rotations for $\mathrm{Q^A}$ are analytically performed during the process. The rise and fall times are again fixed to $\tau_0=100\times 2\pi/\tilde{\omega}_\mathrm{B}=20.5\ \mathrm{ns}$. We find $\tau$ when $\mathcal{C}_0$ and $\mathcal{C}_1$ take their maximum values. We obtain $\tau=169.3\ \mathrm{ns}$ and the maximum $\mathcal{C}=100.00\%$ for both $|{\it{\Phi}}_0\rangle$ and $|{\it{\Phi}}_1\rangle$. Figure \ref{fig:025m}(c) depicts the time evolution during the echoed cross-resonance pulse for each initial state $|{\it{\Phi}}_0\rangle$ and $|{\it{\Phi}}_1\rangle$. 

We have examined the case where the cross-resonance signal is input from a higher-frequency qubit to a lower-frequency qubit. Here, conversely, the qubit frequency detunings are set to ${\it{\Delta}}_\mathrm{A}=-0.3\omega_\mathrm{FSR}$ and ${\it{\Delta}}_\mathrm{B}=0.7\omega_\mathrm{FSR}$. We confirm almost the same results with an echoed cross-resonance pulse, $\mathcal{C}_0=99.99\%$, $\mathcal{C}_1=100.00\%$, and $\tau=160.3$ ns, as those described above. Thus, we continue our study by focusing on a cross-resonance signal from a higher-frequency qubit.

One important consideration for remote cross-resonance gates is the degree of entanglement generation \cite{Niu2023,Zhong2021,Kurpiers2018,Leung2019}. We simulate time evolution with the echoed remote cross-resonance pulse determined above for the initial qubit state of $|{\it{\Phi}}^\mathrm{I}_+\rangle=(|00\rangle+|10\rangle)/\sqrt{2}$. When $[ZX]^{1/2}$ is applied to $|{\it{\Phi}}^\mathrm{I}_+\rangle$, we obtain $|{\it{\Phi}}^\mathrm{F}_+\rangle=(|00\rangle-i|01\rangle+|10\rangle+i|11\rangle)/2$, which is an entangled state, unlike $|{\it{\Phi}}^\mathrm{F}_0\rangle$ and  $|{\it{\Phi}}^\mathrm{F}_1\rangle$. With a qubit state $|{\it{\Phi}}_+\rangle$, the concurrence as a measure of entanglement is defined by
\begin{equation}
\label{concurrence+}
\mathcal{C}_+ = 16|c_{+,|00\rangle}||c_{+,|01\rangle}||c_{+,|10\rangle}||c_{+,|11\rangle}|
\end{equation}
so that this becomes 1 when $|{\it{\Phi}}_+\rangle=|{\it{\Phi}}^\mathrm{F}_+\rangle$. We obtain the concurrence $\mathcal{C}_+=100.00\%$. Thus, we examine the qubit frequency dependence of $\mathcal{C}_+$ in the next section.

\subsection{Qubit frequency dependence}

Let us consider the range of qubit frequencies in which the remote cross-resonance gate operates efficiently. We examine the maximum value of $\mathcal{C}_+$ with an echoed cross-resonance pulse for each $\delta_\mathrm{A}$ and $\delta_\mathrm{B}$ in the ranges $\delta_i \neq 0,1,2\cdots$, where $\delta_i=|{\it{\Delta}}_i|/\omega_\mathrm{FSR}, i=\mathrm{A}$ and $\mathrm{B}$, because $\delta_i=0,1,2,\cdots$ do not well define the qubit state owing to the same energy levels of the qubit and cable mode [Fig. \ref{fig:concept}(b)].

\subsubsection{0.25-m coaxial cable}

\begin{figure}[tbh]
\includegraphics[width=80mm]{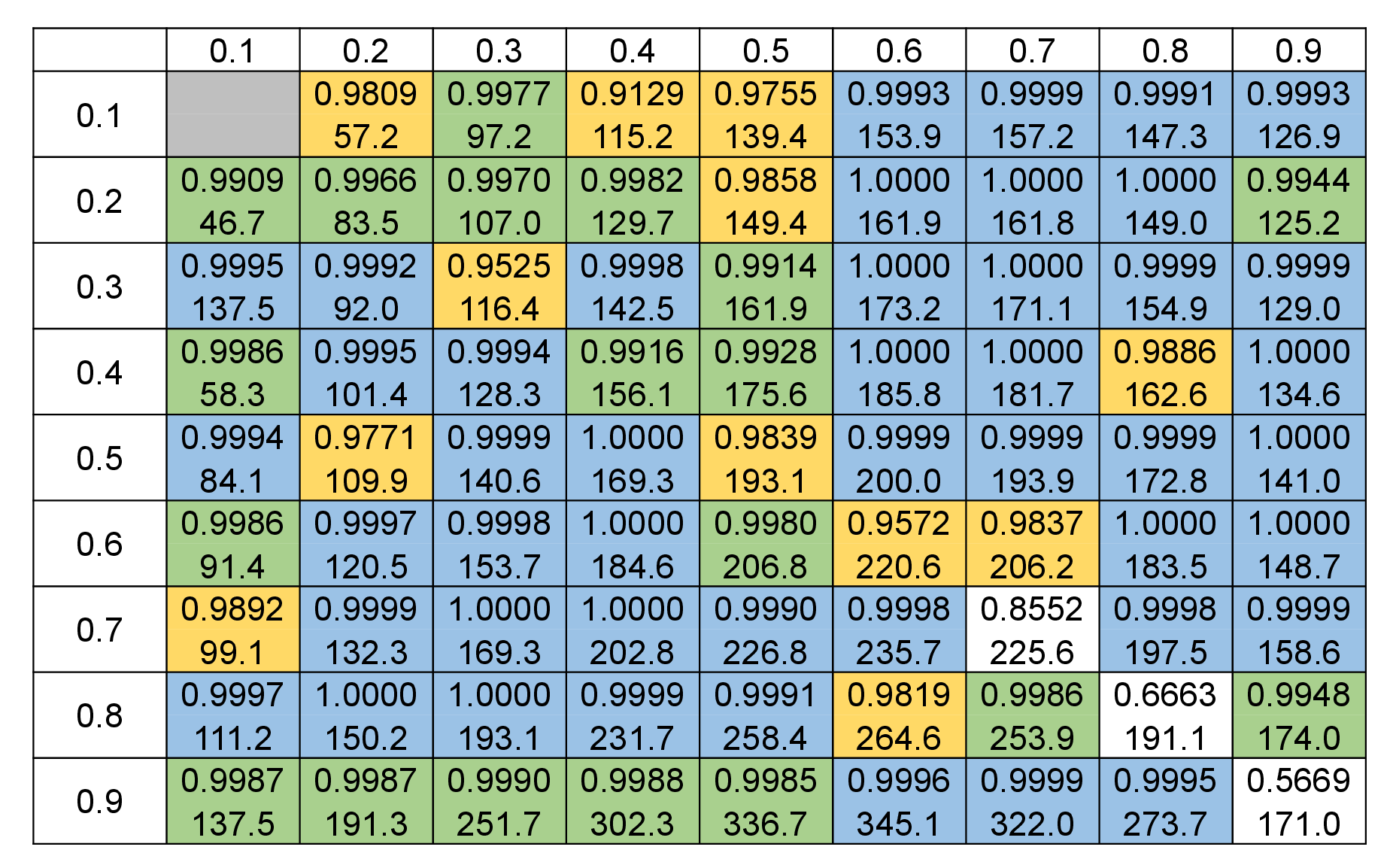}
\centering
\caption{\label{fig:025m_summary}Time evolution simulation results of an echoed cross-resonance pulse with the initial qubit state of $|{\it{\Phi}}^\mathrm{I}_\mathrm{+}\rangle=(|00\rangle+|10\rangle)/\sqrt{2}$ for a 0.25-m coaxial cable with $g/(2\pi)=0.03$ GHz. The numbers in the titles of rows and columns are the qubit frequency detunings $\delta_\mathrm{A}=|{\it{\Delta}}_\mathrm{A}|/\omega_\mathrm{FSR}$ and $\delta_\mathrm{B}=||{\it{\Delta}}_\mathrm{B}|/\omega_\mathrm{FSR}$, respectively.  In each cell, the top figure is the concurrence $\mathcal{C}_+$ and the bottom figure is the duration $\tau$ in ns. The cells are colored according to their $\mathcal{C}_+$ values (blue: $>$99.9\%, green: $>$99\%, orange: $>$90\%, and white: $\leq$90\%).}
\end{figure}

Figure \ref{fig:025m_summary} summarizes the maximum $\mathcal{C}_+$ and echoed pulse duration $\tau$ in the range $0.1 \leq \delta_\mathrm{A} \leq 0.9$ and $0.1 \leq \delta_\mathrm{B} \leq 0.9$ for a 0.25-m coaxial cable with $g/(2\pi) = 0.03\ \mathrm{GHz}$.
We can find $3\times3$ cells with $\mathcal{C}_+>99.9\%$ in the top-right part ($0.1 \leq \delta_\mathrm{A} \leq 0.3$ and $0.6 \leq \delta_\mathrm{B} \leq 0.8$) with $\tau\sim160\ \mathrm{ns}$. The frequency detuning of $\delta_i=0.1$ corresponds to 38.5 MHz.
This means that if we aim for the frequencies of $\delta_\mathrm{A} = 0.2$ and  $\delta_\mathrm{B} = 0.7$ to fabricate the qubits, even with $\pm$38-MHz frequency error due to manufacturing, obtaining $\mathcal{C}_+>99.9\%$ is possible by calibrating the signal frequency and pulse duration. A standard deviation of the frequency distribution of $\sigma_f=14\ \mathrm{MHz}$ has already been achieved with the current technology \cite {Hertzberg2021, Zhang2022}. 

We can find other $3\times3$ cells with $\mathcal{C}_+>99.9\%$ in the bottom left ($0.6 \leq \delta_\mathrm{A} \leq 0.8$ and $0.2 \leq \delta_\mathrm{B} \leq 0.4$). The results for the center of this area, $\delta_\mathrm{A}=0.7$ and $\delta_\mathrm{B}=0.3$, have been shown in the previous section and are summarized in Table \ref{table:average} as an example.
The maximum $\mathcal{C}_+$ in the diagonal cells is smaller than the others. We find that this degradation is due to the closeness of the energy levels of the qubit $|11\rangle$ state and the cable state comprising two cable modes $M\pm k$. This is a notable feature in remote cross-resonance gates through cables with multimode coupling.

A smaller coupling of $g/(2\pi)=0.02$ GHz is also examined (\ref{A:rcr}, Fig. \ref{fig:s_summary}(a)). Despite the increase in $\tau$ ($\sim$300 ns), the range with $\mathcal{C}_+>99.9\%$ does not expand. Although some cells with $\mathcal{C}_+\leq99\%$ change to $\mathcal{C}_+>99\%$ and the cell where the qubit state is not defined disappears, it is not beneficial to make the coupling smaller in this case, where high $\mathcal{C}_+$ values are already obtained over a wide range.

\subsubsection{0.5-m coaxial cable}

\begin{figure}[htb]
\centering
\includegraphics[width=80mm]{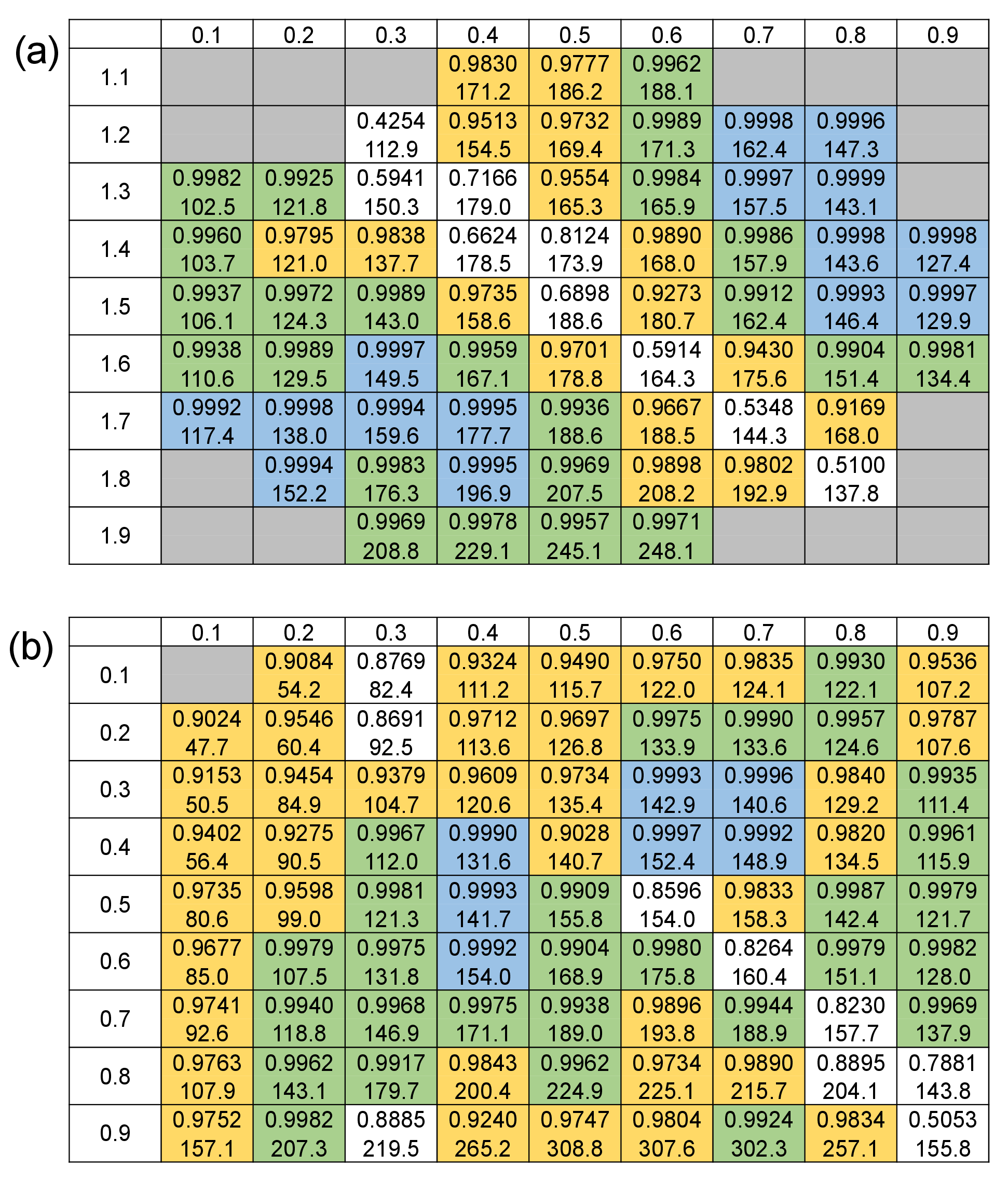}
\caption{\label{fig:05m_summary}Time evolution simulation results of an echoed cross-resonance pulse with the initial qubit state of $|{\it{\Phi}}^\mathrm{I}_\mathrm{+}\rangle=(|00\rangle+|10\rangle)/\sqrt{2}$ for a 0.5-m coaxial cable: (a) $g/(2\pi)=0.03$ GHz and (b) $g/(2\pi)=0.02$ GHz. Other explanations are the same as those in Fig. \ref{fig:025m_summary}}
\end{figure}

The free spectral range for a 0.5-m coaxial cable is a small value of $\omega_\mathrm{FSR}/(2\pi)=0.2174$ GHz (Table \ref{table:path}). We consider the range $1.1\leq\delta_\mathrm{A}\leq1.9$ and $0.1\leq\delta_\mathrm{B}\leq0.9$ to include the region ${\it{\Delta}}/(2\pi)\sim0.4$ GHz where high $\mathcal{C}_+$ is obtained for a 0.25-m coaxial cable. The results for $g/(2\pi)=0.03$ GHz are shown in Fig. \ref{fig:05m_summary}(a). As in the case of a 0.25-m coaxial cable, we can find two $2\times2$ cells with $\mathcal{C}_+>99.9\%$ [($1.2\leq\delta_\mathrm{A}\leq1.3$ and $0.7\leq\delta_\mathrm{B}\leq0.8$) and ($1.4 \leq\delta_\mathrm{A}\leq1.5$ and $0.8\leq\delta_\mathrm{B}\leq0.9$)] for $\tau\sim150$ ns. The frequency detuning of $\delta_i=0.1$ corresponds to 21.7 MHz. This means that if we aim for the frequencies of the center of the areas to fabricate the qubits, obtaining $\mathcal{C}_+>99.9\%$ is possible even with $\pm$10 MHz frequency error owing to manufacturing. The center of the area $\delta_\mathrm{A}=1.25$ and  $\delta_\mathrm{B}=0.75$ gives $\mathcal{C}_+=99.98\%$, which is included in Table \ref{table:average}, and $\delta_\mathrm{A}=1.45$ and $\delta_\mathrm{B}=0.85$ also gives $\mathcal{C}_+=99.98\%$. We find some $2\times2$ cells with $\mathcal{C}_+>99.9\%$ in the range $0.1\leq\delta_\mathrm{A} \leq0.9$ and $1.1\leq\delta_\mathrm{B}\leq 1.9$ (\ref{A:rcr}, Fig. \ref{fig:s_summary}(b)).

The range $0.1 \leq \delta_\mathrm{A} \leq 0.9$ and $0.1 \leq \delta_\mathrm{B} \leq 0.9$, where ${\it{\Delta}}$ exhibits small values, are also examined. As shown in Fig. \ref{fig:05m_summary}(b), reducing the coupling to $g/(2\pi)=0.02$ GHz leads to $2\times2$ cells with $\mathcal{C}_+>99.9\%$ and the average values of $\tau\sim150$ ns and ${\it{\Delta}}/(2\pi)\sim0.22$ GHz. The center of the area $\delta_\mathrm{A} = 0.35$ and  $\delta_\mathrm{B} = 0.65$ gives $\mathcal{C}_+=99.98\%$ for $\tau=147.3$ ns, which is also included in Table \ref{table:average}. 

\subsubsection{1-m coaxial cable}

\begin{figure}[tbh]
\centering
\includegraphics[width=80mm]{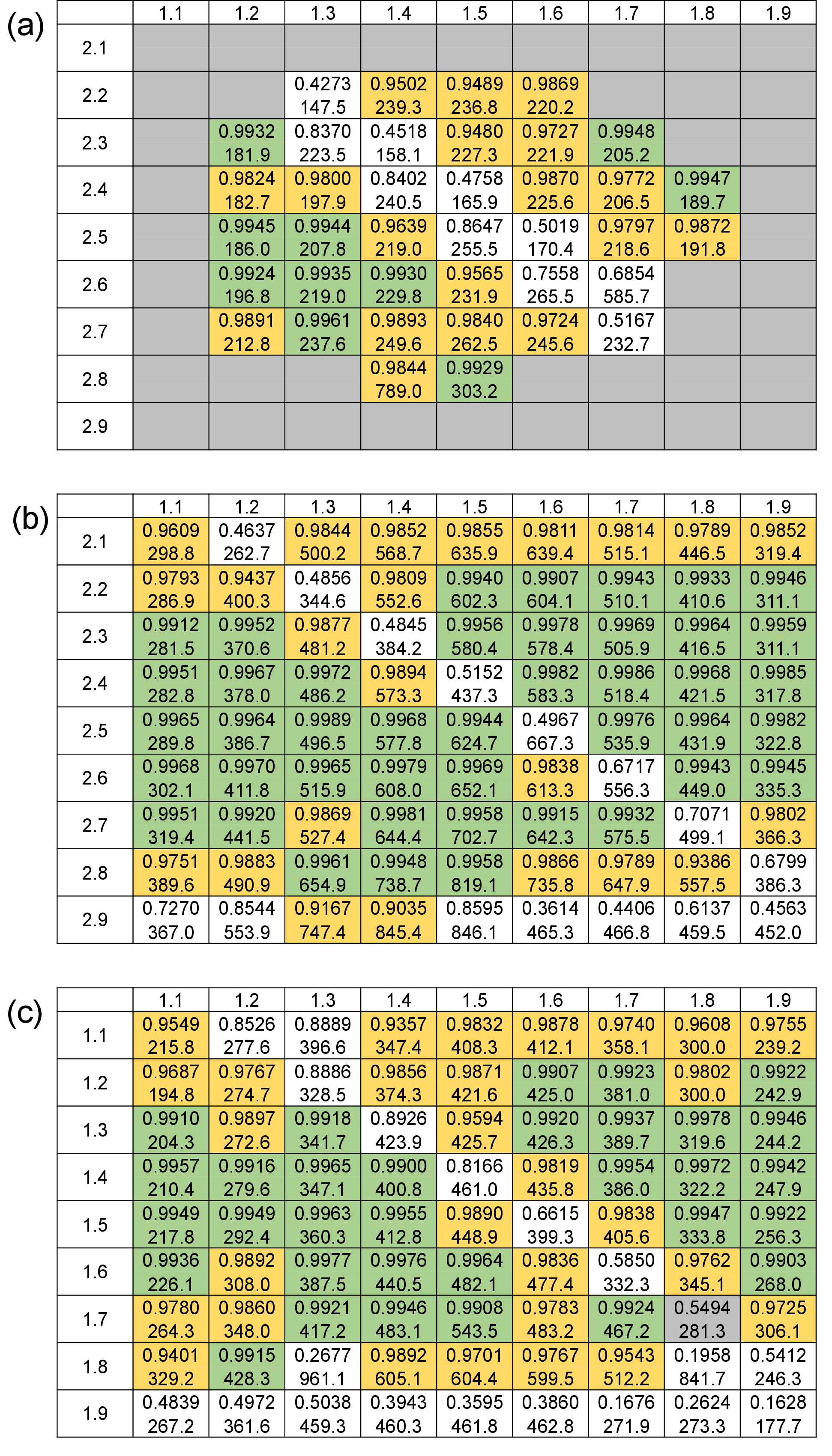}
\caption{\label{fig:1m_summary}Time evolution simulation results of an echoed cross-resonance pulse with the initial qubit state of $|{\it{\Phi}}^\mathrm{I}_\mathrm{+}\rangle=(|00\rangle+|10\rangle)/\sqrt{2}$ for a 1-m coaxial cable: (a) $g/(2\pi)=0.02$ GHz and (b) (c) $g/(2\pi)=0.01$ GHz. Other explanations are the same as those in Fig. \ref{fig:025m_summary}.}
\end{figure}

The free spectral range for a 1-m coaxial cable is a further small value of $\omega_\mathrm{FSR}/(2\pi)=0.1163$ GHz (Table \ref{table:path}). The range $2.1\leq\delta_\mathrm{A} \leq2.9$ and $1.1\leq\delta_\mathrm{B}\leq1.9$ includes the region ${\it{\Delta}}/(2\pi)\sim0.4$ GHz where high $\mathcal{C}_+$ is obtained for a 0.25-m coaxial cable. However, the coupling $g/(2\pi)=0.03$ GHz does not provide well-defined qubit states over the entire range. The results for the range with $g/(2\pi)=0.02$ GHz are presented in Fig. \ref{fig:1m_summary}(a).  
We find $2\times2$ cells with $\mathcal{C}_+>99\%$ in the bottom-left part ($2.5 \leq \delta_\mathrm{A} \leq 2.6$ and $1.2 \leq \delta_\mathrm{B} \leq 1.3$). The center of the area $\delta_\mathrm{A}=2.55$ and $\delta_\mathrm{B}=1.25$ gives $\mathcal{C}_+=99.28\%$, and the results are summarized in Table \ref{table:average}.  We cannot obtain an area with $\mathcal{C}_+>99.9\%$ as in the cases of 0.25- and 0.5-m cables. It is found that the occupation probability of the cable states with energies corresponding to $\sim2\tilde{\omega}_\mathrm{B}$ or $\sim3\tilde{\omega}_\mathrm{B}$ gradually increases during the cross-resonance pulse. This can be the most important characteristic with regard to remote cross-resonance gates through cables with multimode coupling. However, if we aim for the frequencies of $\delta_\mathrm{A}=2.55$ and $\delta_\mathrm{B}=1.25$ to fabricate the qubits, even in case of manufacturing errors of $\pm5\ \mathrm{MHz}$, it is possible to obtain $\mathcal{C}_+>99\%$, which is in sharp contrast to quantum state transfer, where a detuning of 5 MHz substantially reduces the efficiency, as mentioned earlier. The frequency shift of 5 MHz corresponds to 0.1\% of the qubit frequency $\sim$5 GHz and matches the standard deviation of the frequency distribution that is required to achieve a 1000-qubit computer \cite{Hertzberg2021,Zhang2022}. We also find two $2\times2$ cells with $\mathcal{C}_+>99\%$ in the range $1.1 \leq \delta_\mathrm{A} \leq 1.9$ and $2.1 \leq \delta_\mathrm{B} \leq 2.9$ (\ref{A:rcr}, Fig. \ref{fig:s_summary}(c)).

Unlike in the case of a 0.25-m coaxial cable, it is very beneficial to reduce the coupling to $g/(2\pi)=0.01$ GHz. Although $\tau$ increases, we can find much wider areas with $\mathcal{C}_+>99\%$ as shown in Fig. \ref{fig:1m_summary}(b). For example, there are  $3\times3$ cells ($2.4 \leq \delta_\mathrm{A} \leq 2.6$ and $1.1 \leq \delta_\mathrm{B} \leq 1.3$) with $\mathcal{C}_+>99.5\%$ for $\tau\sim400\ \mathrm{ns}$. This corresponds to allowing manufacturing errors of $\pm11\ \mathrm{MHz}$ for $\mathcal{C}_+>99.5\%$. The center of the area $\delta_\mathrm{A}=2.5$ and $\delta_\mathrm{B}=1.2$ gives $\mathcal{C}_+=99.64\%$, which is included in Table \ref{table:average}. 
The range $1.1 \leq \delta_\mathrm{A} \leq 1.9$ and $1.1 \leq 1.9$ for $g/(2\pi)=0.01\ \mathrm{GHz}$ are also examined, and the results are presented in Fig. \ref{fig:1m_summary}(c). Numerous $2\times2$ cells with $\mathcal{C}_+>99\%$ are obtained. Among them, the area centered on $\delta_\mathrm{A}=1.55$ and $\delta_\mathrm{B}=1.35$ leads to $\mathcal{C}_+>99.5\%$. The results on the center of $\mathcal{C}_+=99.84\%$ and $\tau=400.3$ ns are also included in Table \ref{table:average}.

To summarize the results thus far, as shown in Table \ref{table:average}, there are qubit frequencies of $\mathcal{C}_+>99.9\%$ for 0.25- and 0.5-m cables even with manufacturing errors of $\pm$38 MHz and $\pm$10 MHz, respectively, and $\mathcal{C}_+>99\%$ for a 1-m cable even with $\pm$5-MHz. In the  case of a 1-m cable, by reducing the coupling between the qubits and cable, although $\tau$ increases, the lower limit of $\mathcal{C}_+$ and allowed manufacturing error can be 99.5\% and $\pm$11 MHz. These are in contrast to quantum state transfer, where a small detuning of 5 MHz substantially reduces the efficiency (\ref{A:transfer}).
\begin{table*}[tbh]
\caption{\label{table:average}Examples of remote cross-resonace gates with a high concurrence: target cable length, qubit frequency detunings ($\delta_\mathrm{A}$,$\delta_\mathrm{B}$, and ${\it{\Delta}}$), coupling ($g$), pulse duration ($\tau$), concurrence for the initial qubit state $|{\it{\Phi}}^\mathrm{I}_+\rangle$ ($\mathcal{C}_+$), average gate fidelity ($\bar{\mathcal{F}}$), and concurrence with dissipation for the initial qubit state $|{\it{\Phi}}^\mathrm{I}_+\rangle$ ($\mathcal{C}^{\prime}_+$).}
\centering
\begin{tabular}{ccccccccc}
\br
Cable length & $\delta_\mathrm{A}$ & $\delta_\mathrm{B}$ & ${\it{\Delta}}/(2\pi)$ & $g/(2\pi)$ & $\tau$ & $\mathcal{C}_+$ & $\bar{\mathcal{F}}$ & $\mathcal{C}^\prime_+$\\
(m) &  &  & (GHz) & (GHz) & (ns) &  &  & \\
\mr
0.25 & 0.7 & 0.3 & 0.3846 & 0.03 & 169.3 & 1.0000 & 0.9999 & 0.9999\\
0.5 & 1.25 & 0.75 & 0.4348 & 0.03 & 152.4 & 0.9998 & 0.9998 & 0.9997\\
    & 0.35 & 0.65 & 0.2174 & 0.02 & 147.3 & 0.9998 & 0.9998 & 0.9996\\
1.0 & 2.55 & 1.25 & 0.4419 & 0.02 & 201.9 & 0.9928 & 0.9964 & 0.9859\\
     & 2.5 & 1.2 & 0.4302 & 0.01 & 385.9 & 0.9964 & 0.9976 & 0.9943\\
     & 1.55 & 1.35 & 0.3372 & 0.01 & 400.3 & 0.9984 & 0.9989 & 0.9975\\
\br
\end{tabular}
\end{table*}

\subsection{Average gate fidelity}

We have focused on the concurrence as a measure of remote entanglement generation. Herein, we investigate the average gate fidelity as a more general characteristic of two-qubit gates. The average gate fidelity is also important because it is obtained in randomized benchmarking experiments \cite{Marxer2023,Barends2014,Magesan2011,Magesan2012,Sheldon2016B,Harper2017}. We estimate the average gate fidelity $\bar{\mathcal{F}}$ by averaging the final state fidelity $\mathcal{F}_i$ with an echoed remote cross-resonance pulse for sixteen initial qubit states $|{\it{\Phi}}^\mathrm{I}_i\rangle=\{|0\rangle,|1\rangle,|+\rangle, |-\rangle\}\otimes\{|0\rangle,|1\rangle,|+\rangle, |-\rangle\}$, where $|+\rangle=(|0\rangle+|1\rangle)/\sqrt{2}$ and $|-\rangle=(|0\rangle+i|1\rangle)/\sqrt{2}$. The remote cross-resonance gates summarized in Table \ref{table:average} are examined for each coaxial cable length.  The phase of the cross-resonance pulse and the phase rotation for the final qubit states are optimized to maximize $\bar{\mathcal{F}}$. We obtain $\bar{\mathcal{F}}>99.9\%$ for 0.25- and 0.5-m cables and $>99.5\%$ for a 1-m cable, which are also summarized in Table \ref{table:average}. 

\section{Dissipative-transmission path}

We determine the dissipative properties of the transmission path and perform time evolution simulations for remote cross-resonance gates using these dissipative properties.

\subsection{Relaxation time}

\begin{figure}[tbh]
\centering
\includegraphics[width=80mm]{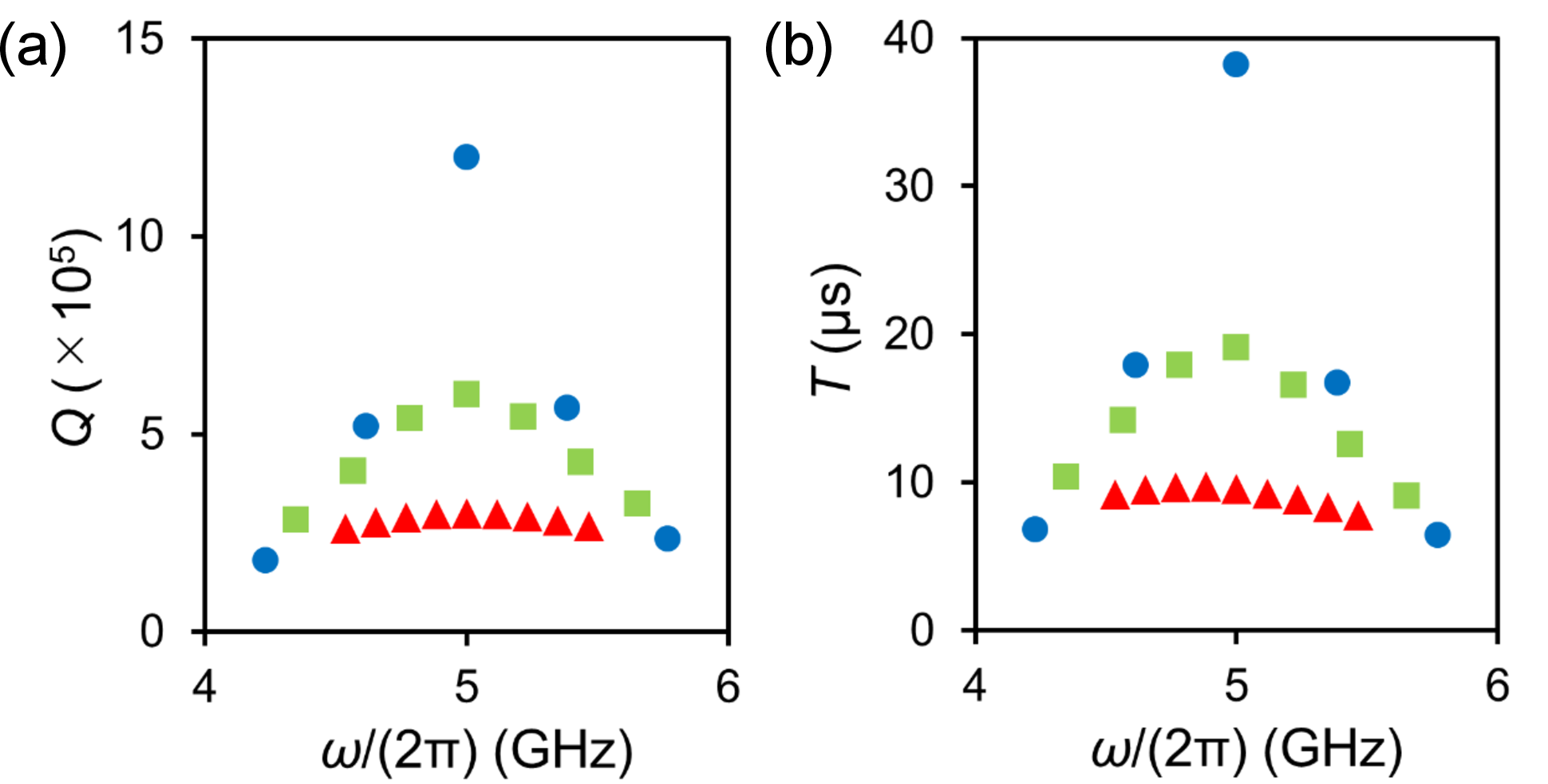}
\caption{\label{fig:cable} (a) $Q$ value and (b) relaxation time $T$ as functions of the frequency $\omega$ of the standing-wave modes of the transmission path. The blue circles, green squares, and red triangles represent the data for 0.25-, 0.5-, and 1-m cables, respectively.}
\end{figure}

The $Q$ value of the $m$th cable standing-wave mode $Q_m$ is given as follows:
\begin{eqnarray}
1/Q_m &=& 1/Q_\mathrm{Cable} + 1/Q_\mathrm{Loss}\nonumber\\
Q_\mathrm{Loss} &=& \frac{\omega_m L}{\mathrm{cos}^2(2\pi l_\mathrm{CPW}/\lambda_\mathrm{CPW}^m)R},
\end{eqnarray}
where $Q_\mathrm{Cable}$ is the cable specific value, which is $1.2\times10^6$ for a 0.25-m coaxial cable \cite{Niu2023}, and $Q_\mathrm{Loss}$ is determined based on the ratio of $l_\mathrm{CPW}$ and $\lambda_\mathrm{CPW}^m$. Here, $l_\mathrm{CPW}$ is presented in Table \ref{table:path} and $\lambda_\mathrm{CPW}^m=2\pi v_\mathrm{CPW}/\omega_m$ is the wavelength of the $m$th mode in the coplanar waveguide, where $v_\mathrm{CPW}=1.157\times10^8$ m/s \cite{Niu2023} and $\omega_m=m\omega_\mathrm{FSR}$. The equivalent inductance $L$ is given by
\begin{equation}
L \approx \frac{1}{2}(2\mathcal{L}_\mathrm{CPW}l_\mathrm{CPW}+\mathcal{L}_\mathrm{Cable} l_\mathrm{Cable}),
\end{equation}
where $\mathcal{L}_\mathrm{CPW}=402$ nH/m and $\mathcal{L}_\mathrm{Cable}=216$ nH/m are the specific inductances of the coplanar wave guide and coaxial cable, respectively \cite{Niu2023}. The resistance between the coplanar waveguide and coaxial cable $R$ is determined to be 0.0749 $\mathrm{n\Omega}$ by reproducing the experimental results \cite{Niu2023}. Figure \ref{fig:cable} shows the $Q$ value and relaxation time of each standing-wave mode for each cable length. The relaxation time of the $m$th mode $T_m$ is obtained by
$Q_m=\omega_m T_m$.

\subsection{Time evolution simulation}

To reveal the effect of dissipation through the multimode transmission paths, we solve the master equations by including the amplitude damping, with the relaxation time $T_m$ determined in the previous section using the QuTip framework \cite{Johansson2012,Johansson2013}. We consider only the dissipation of the transmission path to isolate the cable dissipation effect; in other words, we do not consider decoherence due to qubits. 

We focus on the remote cross-resonance gates presented in Table \ref{table:average} for each cable length. The concurrence for the initial qubit state of $|{\it{\Phi}}_+^\mathrm{I}\rangle$ with dissipation $\mathcal{C}_+^{\prime}$ are examined to compare with $\mathcal{C}_+$. The obtained $\mathcal{C}_+^{\prime}$ are summarized in Table \ref{table:average}. For 0.25- and 0.5-m coaxial cables, $\mathcal{C}_+^{\prime}$ is only 0.01\%--0.02\% less than $\mathcal{C}_+$ and the dissipation effect is negligible. For the 1-m cable, however, the effect is much larger and $\mathcal{C}_+^{\prime}$ is 0.69\% less than $\mathcal{C}_+$ for $g/(2\pi)=0.02$ GHz and 0.09\%--0.21\% for $g/(2\pi)=0.01$ GHz. Reducing the coupling $g$ has an advantage despite the increase in pulse duration $\tau$ even when cable dissipation is considered.

\section{Leakage and single-qubit gate}

Finally, we discuss leakage and single-qubit gates in settings for remote cross-resonance. For the leakage, we examine the time evolution of the initial qubit states of $|00\rangle$, $|01\rangle$, $|10\rangle$, and $|11\rangle$ without any driving pulse, i.e., using the $H_0$ in Eq. (\ref{Hamiltonian}). In view of the results of remote cross-resonance gates, the coupling is set to $g/(2\pi)=0.03$, 0.02, and 0.01 GHz for 0.25-, 0.5-, and 1-m coaxial cables, respectively. The range $0.1\leq\delta_\mathrm{A}\leq0.9$ and $0.1\leq\delta_\mathrm{B}\leq0.9$ is examined for each case. When the dissipation is not considered, the occupation probabilities for all initial qubit states remain 100.00\% in 200 ns for all cable lengths. 

Next, we consider the dissipation of the transmission path. We do not consider the decoherence due to qubits to isolate the cable dissipation effect. Even when the dissipation is considered, the occupation probability of the $|00\rangle$ initial qubit state remains 100.00\% for 200 ns. Other results are presented in \ref{A:leakage}, Figs. \ref{fig:025m_leakage}--\ref{fig:1m_001_leakage}. For 0.25- and 0.5 m cables, 
the occupation probabilities for $|01\rangle$ and $|10\rangle$ in the area corresponding to the remote cross-resonance gates in Table \ref{table:average} are above 99.9\%, while those for $|11\rangle$ are less than 99.9\%.  
For a 1-m cable, the occupation probabilities are between 99.5\% and 99.9\% for all initial qubit states. These results mean that although reducing the coupling leads to considerable fidelities for longer cable lengths, tunable couplers can be useful even when connecting different frequency qubits, as used in previous quantum-state-transfer experiments \cite{Niu2023,Zhong2021,Leung2019,Zhong2019}.

\begin{figure}[tbh]
\centering
\includegraphics[width=80mm]{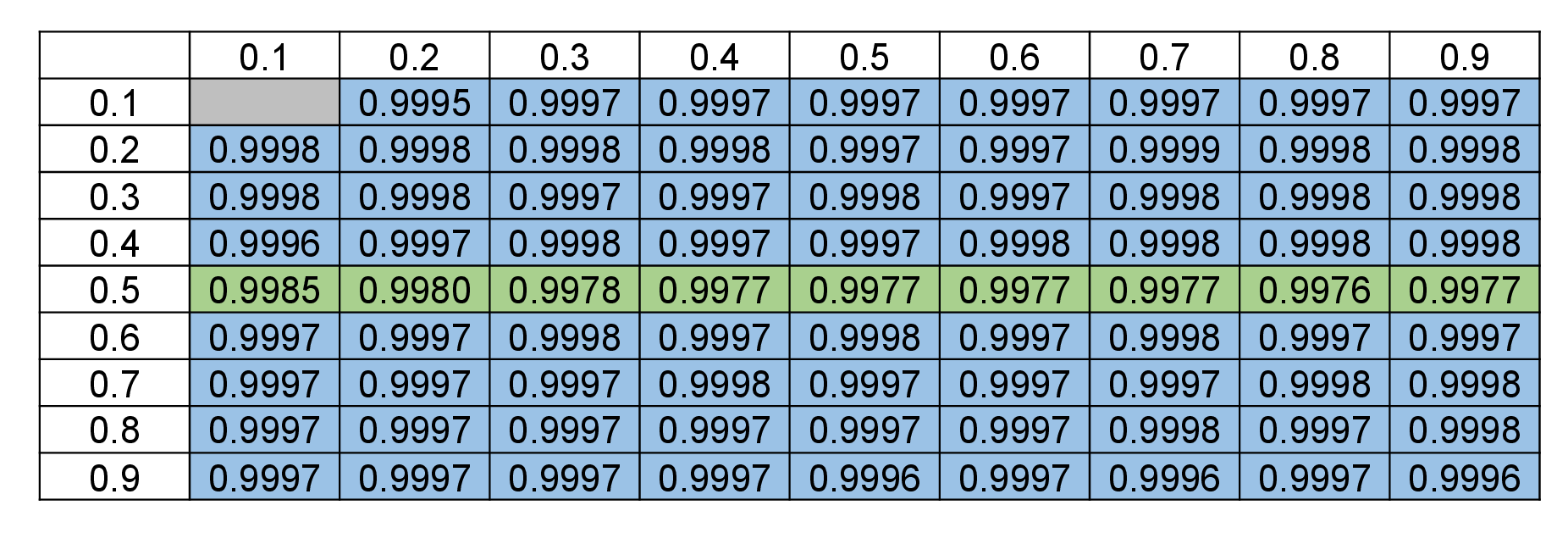}
\caption{\label{fig:025m_one_qubit} Average gate fidelity of $\pi$ rotations around the $x$ axis of $\mathrm{Q^A}$ without dissipation for a 0.25-m coaxial cable. The cells are colored according to their fidelity values. Other explanations are the same as those in Fig. \ref{fig:025m_summary}.}
\end{figure}

For single-qubit gates, we consider a $\pi$ rotation around the $x$ axis of $\mathrm{Q^A}$. In Eq. (\ref{Hamiltonian}), $\tilde{\omega}_\mathrm{B}$ is replaced by $\tilde{\omega}_\mathrm{A}$ and a raised-cosine pulse envelope is used for $\Omega(t)$. The amplitude of the flat part of the signal is set to $\Omega_0=0.005\tilde{\omega}_\mathrm{A}$ and the raised-cosine duration is set to $\tau_0=100\times2\pi/\tilde{\omega}_\mathrm{A}$. The final state fidelities are averaged for the initial qubit states of $|00\rangle$, $|01\rangle$, $|10\rangle$, and $|11\rangle$. The pulse duration is optimized for the maximum averaged state fidelity for each $\delta_\mathrm{A}$ and $\delta_\mathrm{B}$. 
For a 0.25-m cable, the range $0.1\leq\delta_\mathrm{A}\leq0.9$ and $0.1\leq\delta_\mathrm{B}\leq0.9$ is examined. The results are presented in Fig. \ref{fig:025m_one_qubit}.
We find that the fidelities obtained for $\delta_\mathrm{A}=0.5$ are $\sim$99.8\%, which are lower than the others of $>$99.9\%. This is also a notable feature of qubits connected through a cable with multimode coupling. When considering the dissipation of the transmission path, we obtain almost the same results, as shown for remote cross-resonance gates in the previous section. 

For 0.5- and 1.0-m cables, we investigate one parameter set for each length of the remote cross-resonance gates listed in Table \ref{table:average}. The case of $\delta_\mathrm{A}=0.35$, $\delta_\mathrm{B}=0.65$, and $g/(2\pi)=0.02$ GHz for a 0.5-m cable gives average fidelities of 99.99\% and 99.97\% without and with dissipation, respectively. For a 1-m cable, we examine the case of $\delta_\mathrm{A}=1.55$, $\delta_\mathrm{B}=1.35$, and $g/(2\pi)=0.01\ \mathrm{GHz}$. We find that  average fidelities without and with dissipation only go up to 99.19\% and 99.18\%, respectively. In case of a 1-m cable with a narrower mode space, it seems difficult to avoid increasing the occupation probability of states near the energies of $\tilde{\omega}_\mathrm{A}, 2\tilde{\omega}_\mathrm{A},3\tilde{\omega}_\mathrm{A},\cdots$, which comprise the cable modes or their hybridized states with qubit excited states. This means that the use of tunable couplers is increasingly needed.

\section{Conclusion}

We have proposed remote cross-resonance gates to realize high-precision quantum state transfer and remote entanglement between superconducting fixed-frequency qubits. The Hamiltonian for two qubits connected through a multimode cable has been constructed, and time evolution simulations have been performed. For a 0.25-m cable, there are qubit frequencies to exhibit the concurrence in remote entanglement generation $\mathcal{C}_+>99.9\%$ by calibrating the signal frequency and pulse duration, even with $\pm$38-MHz qubit frequency error due to manufacturing. A 0.5-m cable also allows manufacturing errors of $\pm$10 MHz for $\mathcal{C}_+>99.9\%$. We have also found that the average gate fidelity, which is a more general characteristic of two-qubit gates, is higher than 99.9\%. These high-precision quantum interconnects are promising not only for scaling up quantum computer systems but also for nonlocal connections on a chip and may open new avenues such as non-2D quantum error-correcting codes.

For a 1-m cable with a narrower mode spacing, remote cross-resonance gates with $\mathcal{C}_+>99.5\%$ can be achieved by reducing the coupling between the qubits and cable. This is in sharp contrast to quantum state transfer, where a small qubit frequency detuning of 5 MHz notably decreases the efficiency. Although the leakage shows considerable fidelities, it seems difficult to avoid reduction in the fidelity for single-qubit gates on qubits connected a 1-m cable.

The proposed remote cross-resonance gates should first be experimentally tested.  In this context, we must introduce the accuracy improvement methods already implemented in single-qubit gates and cross-resonance gates within chips. It is also necessary to seriously consider utilization of tunable couplers betwee qubits and a coaxial cable even with frequency detunings between the cable mode and the qubits.

\ack
We would like to thank members of the Quantum Hardware project in Quantum Laboratory for their daily knowledge sharing and discussions.

\clearpage

\appendix

\section{\label{A:transfer}Quantum state transfer between frequency-detuned qubits}

\begin{figure}[tbh]
\centering
\includegraphics[width=80mm]{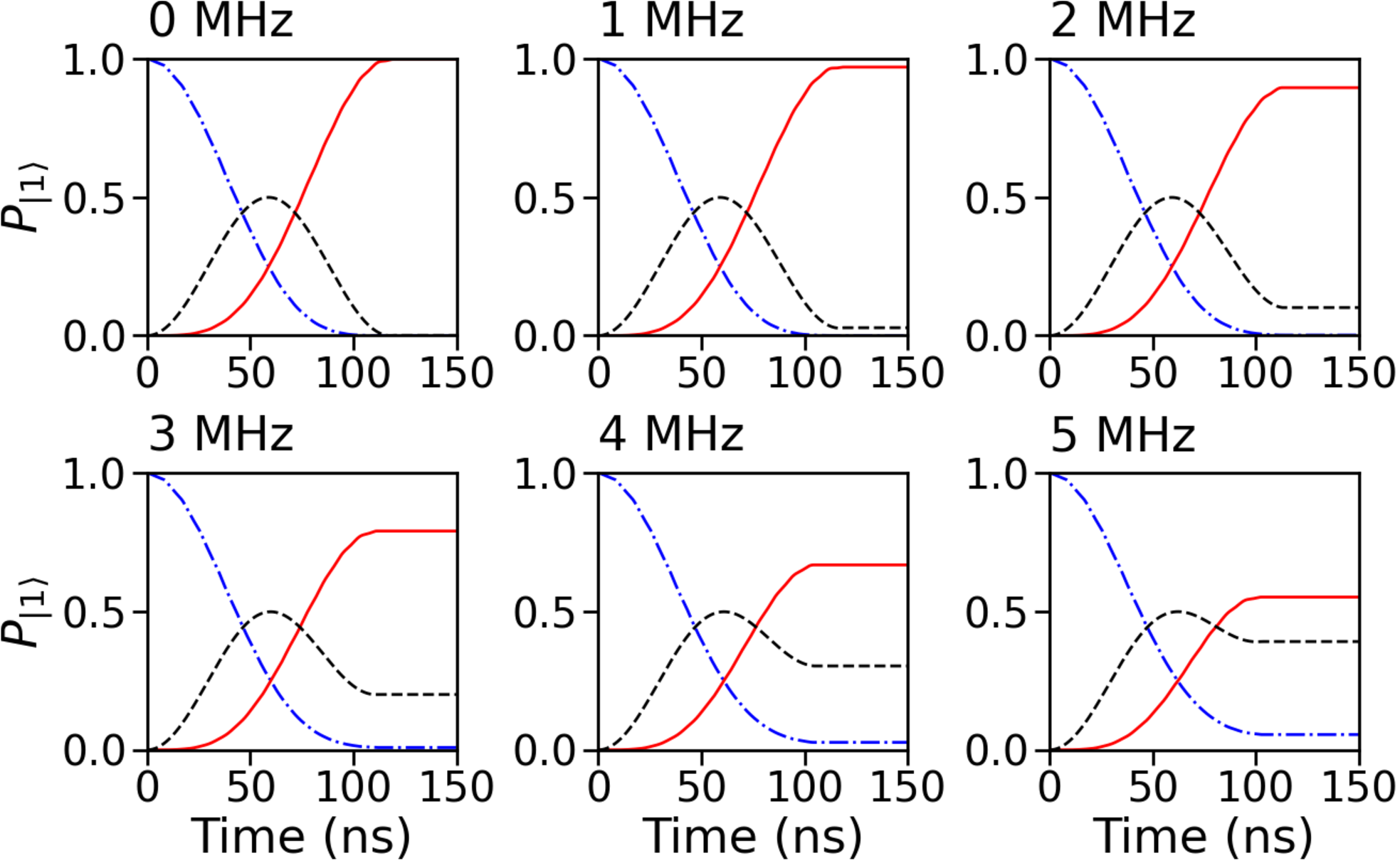}
\caption{\label{fig:detuning}Time evolution of quantum state transfer between frequency-detuned qubits. The qubit frequency detuning of ${\it{\Delta}}/(2\pi)$ is given on the top of each panel. The blue dashed?dotted, red solid, and black dashed curves represent the occupation probabilities of state $|1\rangle$, $P_{|1\rangle}$ for $\mathrm{Q^A}$, $\mathrm{Q^B}$, and the $M$th cable mode, respectively.}
\end{figure}

We show the effect of a qubit frequency detuning ${\it{\Delta}}$ on quantum state transfer via the standing-wave modes of a cable. Two qubits  $\mathrm{Q^A}$ and  $\mathrm{Q^B}$ are connected using a 1-m cable through tunable couplers. Time evolution simulations are performed for the $H_0$ in Eq. (\ref{Hamiltonian}) with $M=43$, $M_\mathrm{Min}=41$, $M_\mathrm{Max}=45$, $\omega_\mathrm{FSR}/(2\pi)=0.1163$ GHz, and $g_\mathrm{A}/(2\pi)=g_\mathrm{B}/(2\pi)=0.003$ GHz. 
After preparing the excited state $|1\rangle$ for $\mathrm{Q^A}$, the tunable couplers are turned on until the occupation probability of state $|1\rangle$, $P_{|1\rangle}$ of $\mathrm{Q^B}$ reaches its maximum. 
The time evolution of $P_{|1\rangle}$ for each qubit and the $M$th cable mode is depicted in Fig. \ref{fig:detuning}.  The final $P_{|1\rangle}$ of $\mathrm{Q^B}$ corresponds to the transfer efficiency. 
The qubit frequency detunings are set to ${\it{\Delta}}_\mathrm{A}/(2\pi)={\it{\Delta}}/(4\pi)+(2\ \mathrm{MHz})$ and ${\it{\Delta}}_\mathrm{B}/(2\pi)=-{\it{\Delta}}/(4\pi)+(2\ \mathrm{MHz})$ for the maximum transfer efficiency. 
We find that a small detuning of ${\it{\Delta}}/(2\pi)=5$ MHz considerably reduces the e?ciency of quantum state transfer.

\section{\label{A:rcr}Qubit frequency dependence of remote cross-resonance gates}

\begin{figure}[tbh]
\centering
\includegraphics[width=80mm]{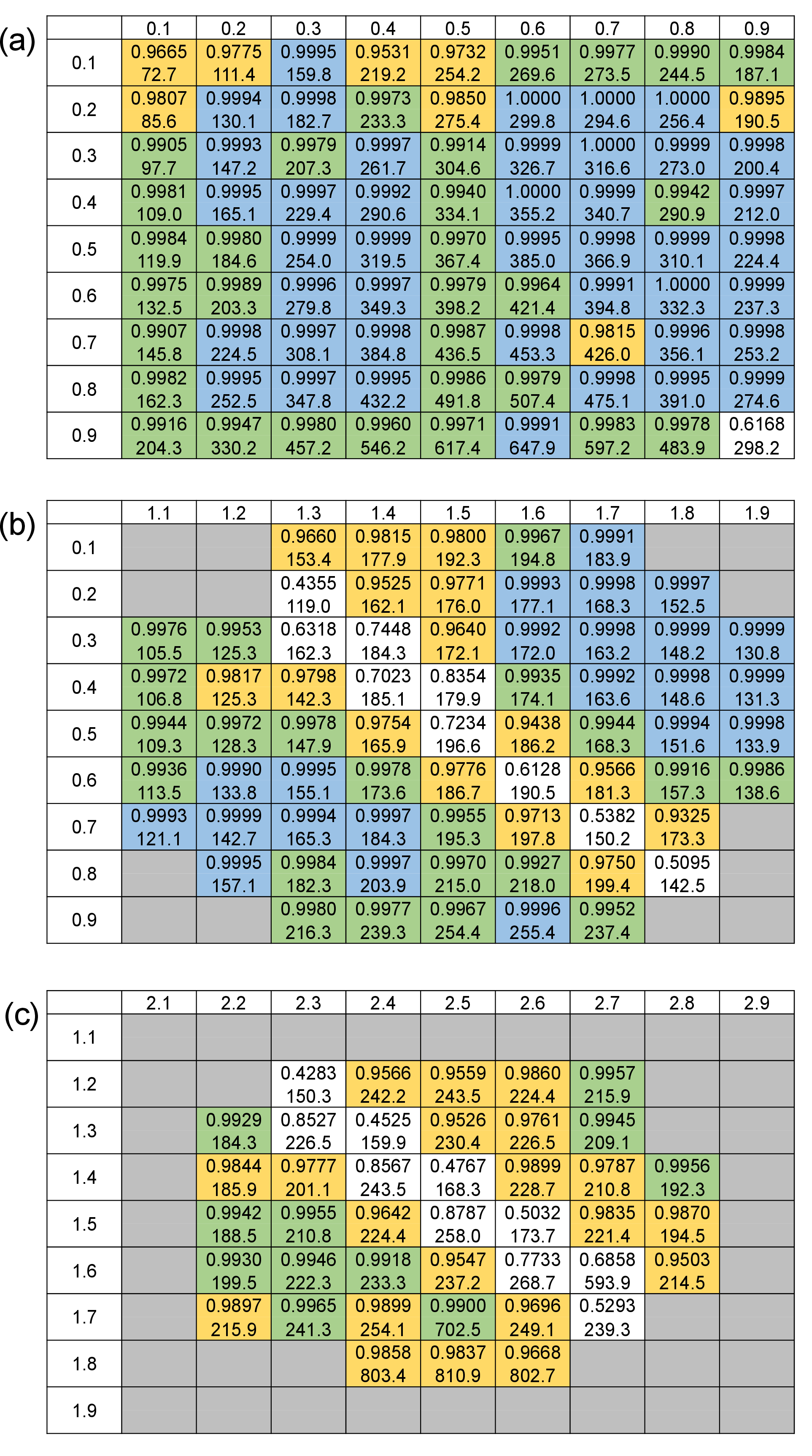}
\caption{\label{fig:s_summary}Time revolution simulation results of an echoed cross-resonance pulse with the initial qubit state of $|{\it{\Phi}}^\mathrm{I}_\mathrm{+}\rangle=(|00\rangle+|10\rangle)/\sqrt{2}$ for a  coaxial cable. The cable length and coupling $g/(2\pi)$ are (a) 0.25 m and 0.02 GHz, (b) 0.5 m and 0.03 GHz, and (c) 1 m  and 0.02 GHz. Other explanations are the same as those in Fig. \ref{fig:025m_summary}.} 
\end{figure}

The primary results for qubit frequency dependence of remote cross-resonance gates are presented in the main text. Here are some additional results.  Figure \ref{fig:s_summary}(a) shows the results of $g/(2\pi)=0.02$ GHz for a 0.25-m cable. We cannot find $3\times3$ cells with $\mathcal{C}_+>99.9\%$ as shown in Fig. \ref{fig:025m_summary}. Figure \ref{fig:s_summary}(b) shows the results of $g/(2\pi)=0.03$ GHz for a 0.5-m cable. We find six $2\times2$ cells with $\mathcal{C}_+>99.9\%$. 
Figures \ref{fig:s_summary}(c) presents the results of $g/(2\pi)=0.02$ GHz for a 1-m cable. We find $2\times2$ cells with $\mathcal{C}_+>99\%$.

\clearpage

\section{\label{A:leakage}Leakage properties}

\begin{figure}[tbh]
\centering
\includegraphics[width=80mm]{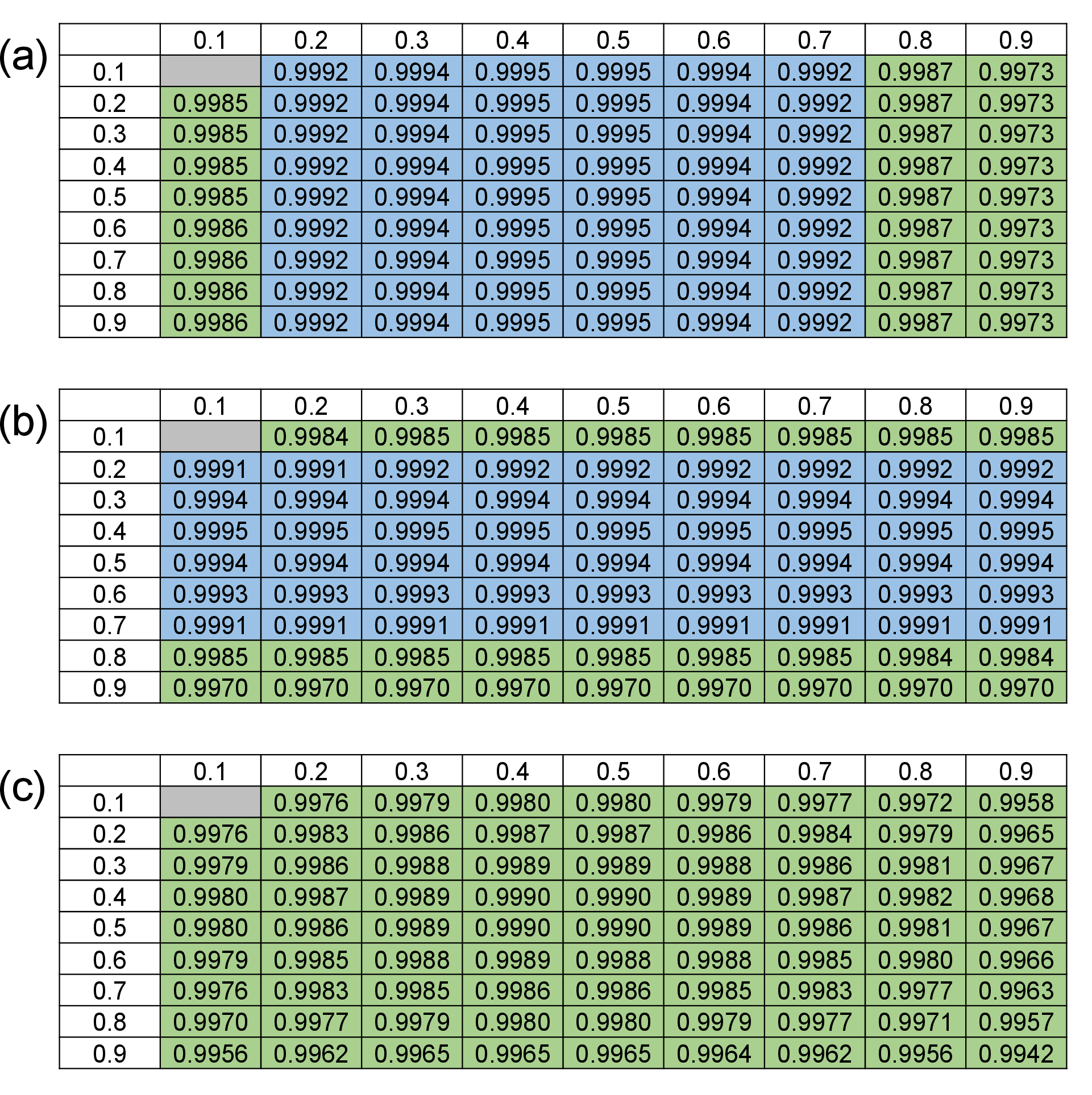}
\caption{\label{fig:025m_leakage}Dissipative leakage properties for a 0.25-m coaxial cable with $g/(2\pi)=0.03\ \mathrm{GHz}$: probabilities that the initial qubit states of (a) $|01\rangle$, (b) $|10\rangle$, and (c) $|11\rangle$ are retained in 200 ns. Other explanations are the same as those in Fig. \ref{fig:025m_summary}.} 
\end{figure}

\begin{figure}[tbh]
\centering
\includegraphics[width=80mm]{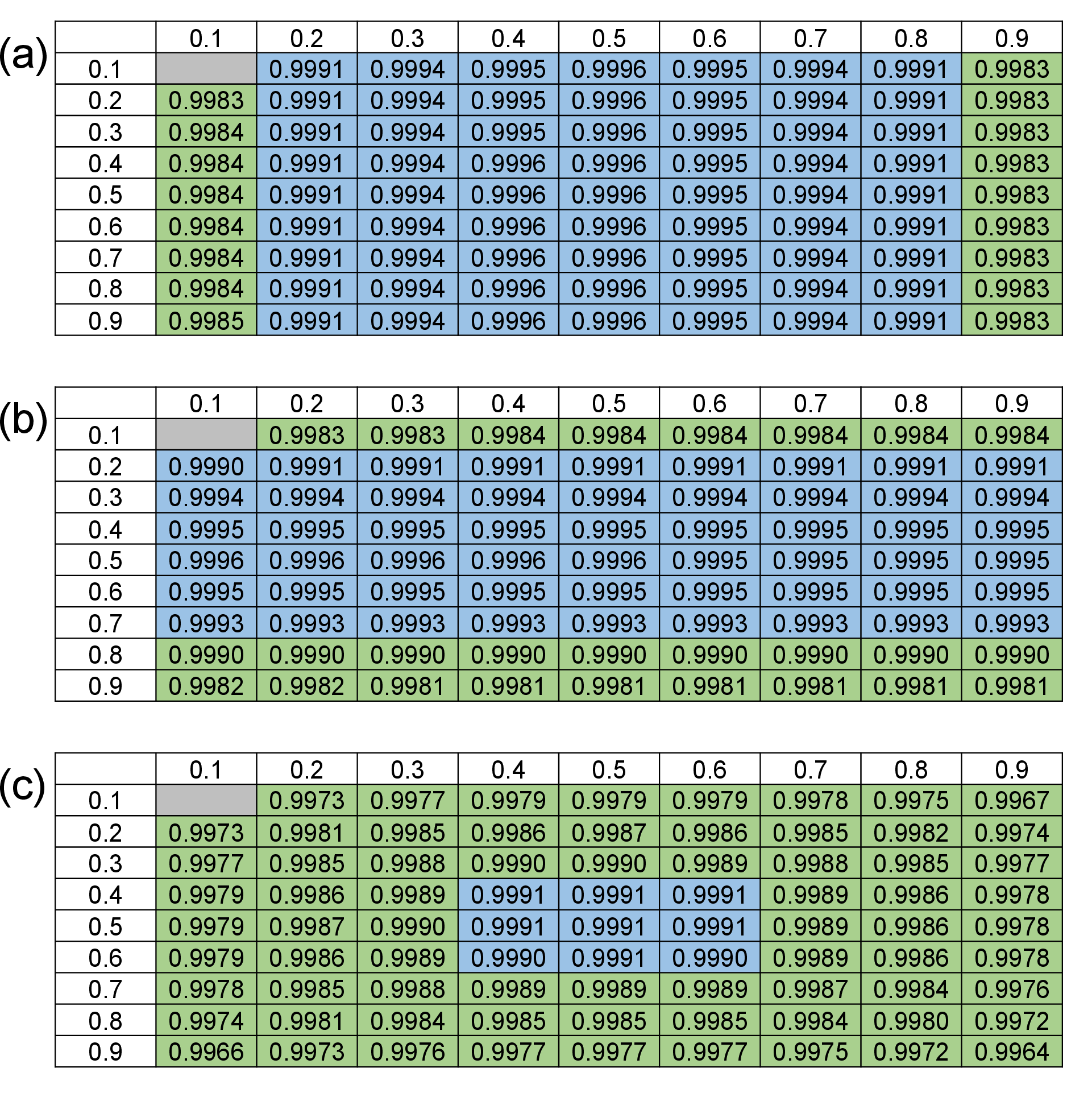}
\caption{\label{fig:05m_002_leakage}Dissipative leakage properties for a 0.5-m coaxial cable with $g/(2\pi)=0.02\ \mathrm{GHz}$: probabilities that the initial qubit states of (a) $|01\rangle$, (b) $|10\rangle$, and (c) $|11\rangle$ are retained in 200 ns. Other explanations are the same as those in Fig. \ref{fig:025m_summary}.} 
\end{figure}

\begin{figure}[tbh]
\centering
\includegraphics[width=80mm]{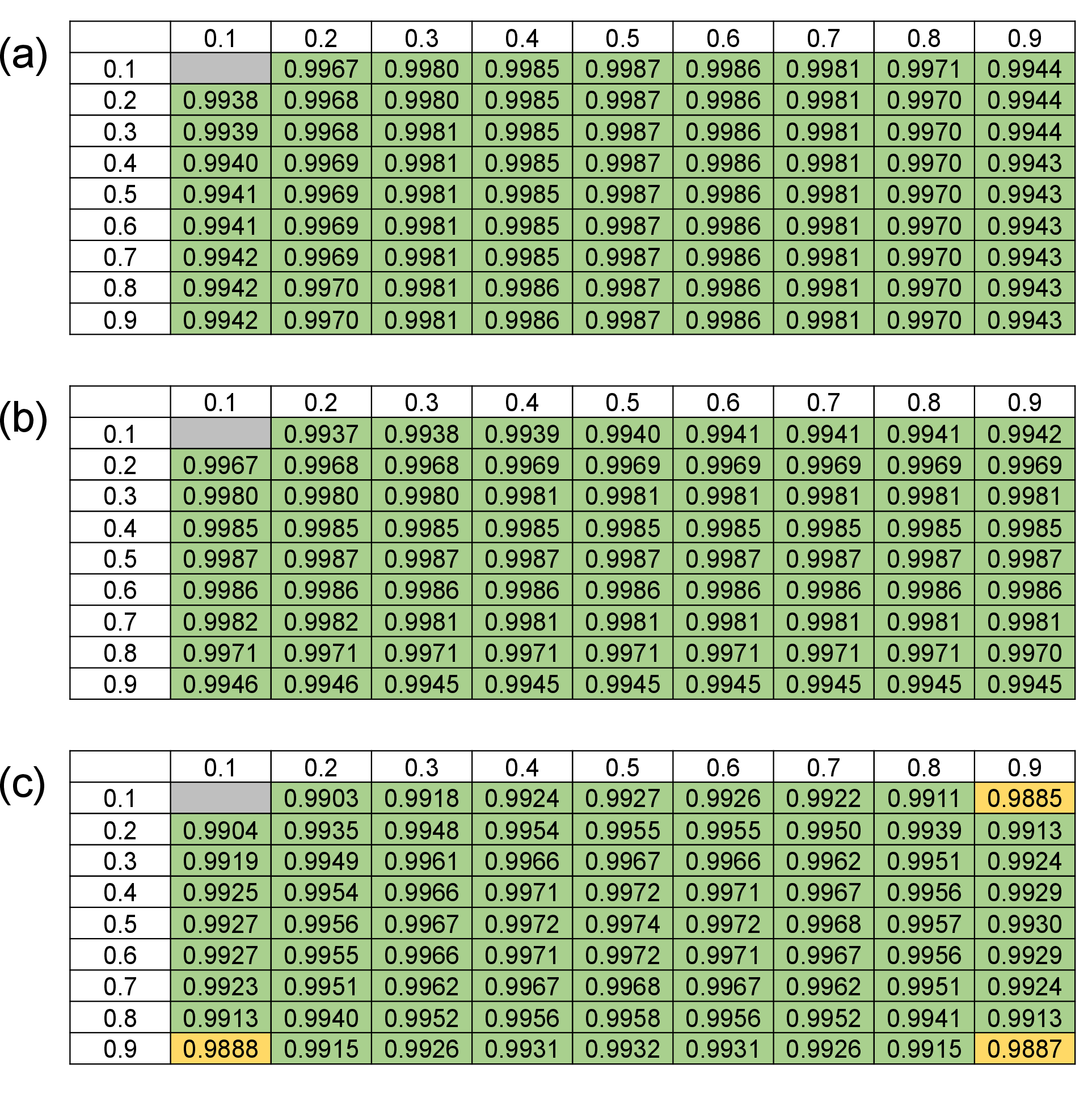}
\caption{\label{fig:1m_001_leakage}Dissipative leakage properties for a 1-m coaxial cable with $g/(2\pi)=0.01\ \mathrm{GHz}$: probabilities that the initial qubit states of (a) $|01\rangle$, (b) $|10\rangle$, and (c) $|11\rangle$ are retained in 200 ns. Other explanations are the same as those in Fig. \ref{fig:025m_summary}.} 
\end{figure}

We examine the time evolution of the initial states of $|00\rangle$, $|01\rangle$, $|10\rangle$, and $|11\rangle$ for 200 ns without any driving pulse, i.e., using the $H_0$ in Eq. (\ref{Hamiltonian}) for the leakage properties. The coupling is set to $g/(2\pi)=0.03, 0.02$, and $0.01\ \mathrm{GHz}$ for 0.25-, 0.5-, and 1-m coaxial cables, respectively. The range $0.1\leq\delta_\mathrm{A}\leq0.9$ and $0.1\leq\delta_\mathrm{B}\leq0.9$ are examined for each case.  We present the results for the initial states of $|01\rangle$, $|10\rangle$, and $|11\rangle$ with cable dissipation in Figs. \ref{fig:025m_leakage}--\ref{fig:1m_001_leakage} because the occupation probabilities without dissipation for all initial qubit states and those with dissipation for the $|00\rangle$ initial qubit state remain 100.00\% for 200 ns. For 0.25- and 0.5 m cables, the occupation probabilities for $|01\rangle$ and $|10\rangle$ in the area corresponding to the remote cross-resonance gates in Table \ref{table:average} are above 99.9\%, while those for $|11\rangle$ are less than 99.9\%.  
For a 1-m cable, the occupation probabilities are between 99.5\% and 99.9\% for all initial qubit states. These results mean that reducing the coupling leads to considerable fidelities for longer cable lengths. 

\clearpage

\bibliography{qst}
\bibliographystyle{iopart-num}

\end{document}